\documentclass{aa}

\usepackage{psfig}

\def\etal{{et\,al.}}

\def\degs{\ifmmode ^{\circ}\else$^{\circ}$\fi}
\def\amin{\ifmmode ^{\prime}\else$^{\prime}$\fi}
\def\asec{\ifmmode ^{\prime\prime}\else$^{\prime\prime}$\fi}
\newbox\grsign \setbox\grsign=\hbox{$>$}
\newdimen\grdimen \grdimen=\ht\grsign
\newbox\laxbox \newbox\gaxbox
\setbox\gaxbox=\hbox{\raise.5ex\hbox{$>$}\llap
     {\lower.5ex\hbox{$\sim$}}}\ht1=\grdimen\dp1=0pt
\setbox\laxbox=\hbox{\raise.5ex\hbox{$<$}\llap
     {\lower.5ex\hbox{$\sim$}}}\ht2=\grdimen\dp2=0pt
\def\gax{\mathrel{\copy\gaxbox}}
\def\lax{\mathrel{\copy\laxbox}}

\def\lsi{LSI +61$\degr$303}

\begin{document}

%   \thesaurus{06         % A&A Section 6: Form. struct. and evolut. of stars
%              (02.01.2;  % Accretion, accretion disks
%               08.02.1:  % binaries: close
%               08.02.3;  % binaries: general
%               08.09.2:  % Stars: individual:
%%               09.18.1:  % (ISM:) reflection nebulae
%               13.25.5)} % X-rays: stars

   \title{The X-ray spectrum of LSI+61$\degr$303}

   \titlerunning{}

   \author{J. Greiner, A. Rau}

   \offprints{J. Greiner, jgreiner@aip.de}

   \institute{Astrophysical Institute
             Potsdam, An der Sternwarte 16, D-14482 Potsdam, Germany
             }

   \date{Received 4 April 2001; accepted ??}

 \abstract{
It had been proposed earlier that the hard X-ray and gamma-ray
radiation of the Be/X-ray system \lsi\ could be due to
inverse Compton scattering of optical photons from the Be star
by the same electron population which also produces the radio emission. 
Recently, Apparao (2001) has calculated this inverse Compton emission
in more detail, and predicted that the X-ray spectrum should show
a break at around 20 keV. We investigated archival RXTE data, but do not
find such a break in the 2--25 keV range. 
%ROSAT data also exclude such a break in the 0.1--2.4 keV range.
The implications of this finding are shortly discussed.
         \keywords{X-ray: stars  -- binaries: close --
                   Stars: individual: \lsi\ $\equiv$ V615 Cas
               }}

   \maketitle

\section{Introduction}

\lsi\ $\equiv$ V615 Cas is a Be binary system which exhibits 
radio outbursts at regular 26.5 day 
intervals which is believed to be the orbital period (Gregory \& Taylor 1978,
Taylor \& Gregory 1982). This source has created particular excitement
because of the possible association with the strong, 100 MeV gamma-ray source
2CG 135+01 (Bignami \& Hermsen 1983). In fact, \lsi\ was discovered
in X-rays while searching for the counterpart of 2CG 135+01 
(Bignami \etal\ 1981). 
Confusion was initially introduced by the
presence of the quasar QSO 0241+622 which is less than 1\fdg5 away from \lsi\
and which contributes to the (blended) emission as measured by OSSE and 
COMPTEL.
In the past years, ROSAT, ASCA and RXTE observations have resolved
the emission from the Be star and the quasar. Moreover, the latest
EGRET analysis has reduced the error circle of 
2CG 135+01 $\equiv$ 2EG J0241+6119 $\equiv$ 3EG J0241+6103 to 11\amin,
thus excluding the quasar as a counterpart candidate as well as making the
association with \lsi\ (which is 13\amin\ apart) quite uncertain 
(Kniffen \etal\ 1997, Hartman \etal\ 1999)!

Nevertheless, the X-ray emission mechanism of \lsi\ has inspired both,
theoreticians as well as observers. While there is general agreement
that the radio outbursts are produced by synchrotron radiation, it has
been argued that it is unlikely that the X-ray emission stems from the
same electron population, but rather could originate by inverse Compton 
emission (Taylor \etal\ 1996, Harrison \etal\ 2000).
This picture is supported by extensive multi-wavelength observations
which are reported in Strickman \etal\ (1998) and Harrison \etal\ (2000).

Very recently, Apparao (2001) has modelled this inverse Compton emission
in more detail. Based on the assumption that the electron population
is rather steady (that is, ignoring the $\sim$0.5 phase shift between
X-ray and radio peak), he derived the properties of this electron population
(such as spectral index, magnetic field and drift velocity) from the
observed radio peak flux, and from that calculates the 10--200 keV spectrum. 
While Apparao (2001) finds general agreement of his model with the data, 
he also predicts a clear spectral break caused by the peak of the optical
photon field (from the Be star) around $\epsilon \sim 10$ eV and boosted by 
the Lorentz factor 
\begin{equation}
 \gamma_c = {4 \pi m c^2 v_{\rm d} r \over L \sigma }
\end{equation}

\noindent
For the values he used, Apparao (2001) derives $\gamma_c \sim 43$,
and thus a break at $\sim \gamma_c^2 \epsilon \sim 20$ keV.
The inverse Compton photon spectrum is therefore given by
(see thick line is Fig. \ref{broadsp})

$$ 1.3 \times 10^{-1} E^{-1.6} ph\, cm^{-2} s^{-1} keV^{-1} {\rm ~~for~ energies~} E \lax 20~ {\rm keV} $$
and
$$ 1.8 \times 10^{-2} E^{-2.1} ph\, cm^{-2} s^{-1} keV^{-1} {\rm ~~for~ energies~} E \gax 20~ {\rm keV} $$

In this short note we have investigated pointed {\it Rossi} X-ray Timing 
Explorer (RXTE; Swank 1998) data to test
this clear prediction for the spectral shape of the X-ray emission. 
Note that while Harrison \etal\ (2000) give a complete
description of these same RXTE data with respect to photon flux and
temporal variability, no clear statement is made about the spectral slope.

\begin{table*}
\caption{Log of RXTE observations. 
%The radio phase refers to
% T$_{0}$=JD 2443366.775 and P$_{orb}$=26.4917 days (Taylor \& Gregory 1984,
% Gregory \etal\ 1999).
}
\vspace{-0.2cm}
\begin{tabular}{cccccccc}
\hline
  ID & Date   & $N_{\rm H}$ & photon & Norm & PCA count & 
    Flux (2--10 keV) & $\chi^{2}$ / dof\\
 (10172-)& (UTC, 1996-)  & (10$^{22}$ cm$^{-2}$) & index & 
   (10$^{-3}$ ph keV$^{-1}$ cm$^{-2}$s$^{-1}$)  & rate (cts/s) & 
   (10$^{-12}$ erg cm$^{-2}$ s$^{-1}$) & \\
\hline
01-01-00 & 03-01 13:41-19:36 &  3.4 & -2.1 & 7.4 & 7.4 & 12.2 & 45/61\\
02-01-00 & 03-04 21:24-01:47 &  5.2 & -2.1 & 9.2 & 8.7 & 13.5 & 39/61\\
03-01-01 & 03-08 00:01-03:59 &  5.2 & -2.3 & 7.9 & 5.3 & 8.5 & 38/61\\
04-01-00 & 03-10 00:55-05:25 &  6.5 & -2.4 & 9.1 & 5.2 & 8.2 & 28/61\\
05-01-00 & 03-13 03:51-08:56 &  3.8 & -2.0 & 5.7 & 7.7 & 11.7 & 48/61\\
06-01-00 & 03-16 00:24-03:08 &  4.8 & -2.1 & 9.7 & 9.7 & 15.2 & 25/61\\
06-01-01 & 03-16 03:39-06:21 &  3.4 & -2.0 & 7.0 & 9.7 & 14.9 & 47/61\\
07-01-00 & 03-18 10:28-15:54 &  5.0 & -2.1 & 14.0 & 13.5 & 21.0 & 61/61\\
09-01-00 & 03-24 23:07-04:35 &  2.1 & -2.3 & 8.0 & 6.5  & 11.6 & 30/55\\
10-01-00 & 03-26 00:52-07:26 &  0.0 & -1.6 & 2.4 & 11   & 11.0 & 43/55\\
11-01-00 & 03-30 03:59-10:41 &  3.4 & -2.1 & 5.2 & 4.8  & 8.4 & 30/55\\
\hline
\end{tabular}
\label{fitres}
\end{table*}

\section{Observations}

We have analyzed 11 pointed RXTE observations of \lsi\ taken between March 1 
and March 30, 1996 (see Tab. \ref{fitres} for a log of all observations). 
All data have been retrieved from the HEASARC archive. 
Similar to Harrison \etal\ (2000) we have used ``Standard 2''
mode spectra of the Proportional Counter Array (PCA; Jahoda \etal\ 1996) 
which have full energy resolution and  16-sec accumulation time.
The only major difference in our treatment as compared to that of 
Harrison \etal\ (2000) is that we do not only use data from layer 1 of 
the PCA units (PCUs).
Instead, we use all layers, since it is predominantly in the lower layers
where the harder ($>$10 keV) photons are detected.

Unfortunately, the observations were effected by several instrumental 
problems, and the PCA gain had to be changed (see Harrison 
\etal\ 2000 for a complete description). Therefore,
two different sets of background models (faintl7\_e01v03 
and faint240\_e01v03 for data from epoch 1 (until March 20) 
and faintl7\_e2v19990824 and faint240\_e2v19990909 
for data from epoch 2 (from March 21 on)) had to be used.
All 5 PCUs were used except for March 24, when
only 3 PCUs of the PCA were available.
We combined all 16-sec spectra into one X-ray spectrum per observation. 
For spectral analysis within the XSPEC package (Arnaud 1996), we used a model 
consisting of cold absorption and a power law model. 
The fit is acceptable in all cases, and the detailed fit results are shown 
in Table \ref{fitres}.

\begin{figure}[bh]
 \vbox{\psfig{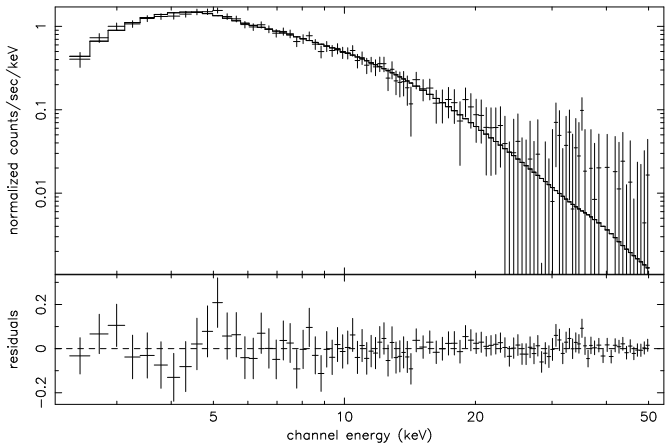}}\par
% im gedrehten Bild  oben      links      unten
\vspace{-0.2cm}
\caption[fc]{PCA X-ray spectrum of \lsi\ from March 18, 1996. Upper panel: 
  count rate spectrum and fitted model, consisting of cold absorption and a 
  power law model. Lower panel: residuals in units of cts/sec/keV. 
  Note the feature near 5~keV.}
\label{sample}
\end{figure}

As an example, the spectrum of March 18, without any further binning, 
is shown in Fig. \ref{sample}.
As clearly visible, the spectrum gets noisy above $\sim$25 keV at which point
the steeper source spectrum approaches the intensity of the background.
Therefore, we limited our spectral fitting to the 2.3--25~keV range.
The positive residuals around 35 keV suggest that the flux above $\sim$30 keV
is mostly background, and therefore the intrinsic spectrum of \lsi\ is
steeper at these energies. Note that no useful data are accumulated in the
HEXTE clusters (as mentioned already by Harrison \etal\ 2000).

\begin{figure}[th]
\vbox{\psfig{figure=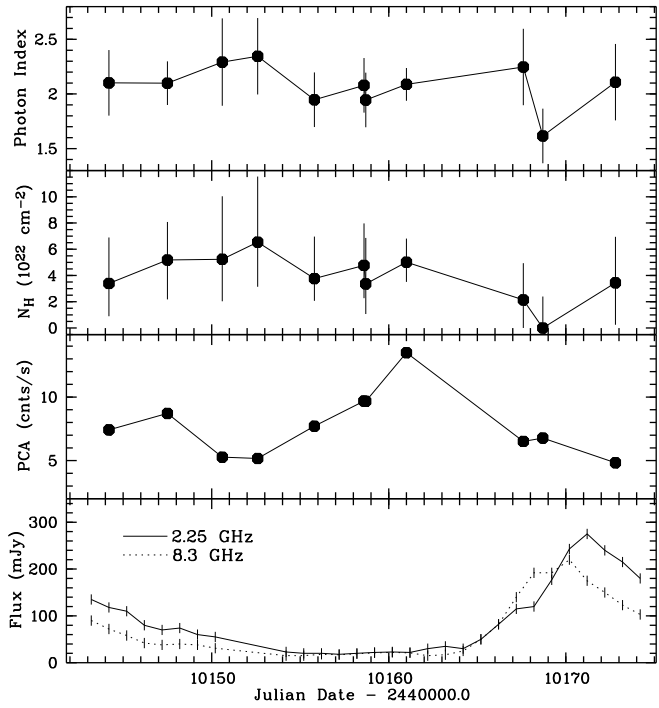,width=8.8cm,%
      bbllx=2.3cm,bblly=7.8cm,bburx=18.7cm,bbury=25.7cm,clip=}}\par
% im gedrehten Bild  oben      links      unten
\vspace{-0.2cm}
\caption[xpar]{Variation of the X-ray spectral fit parameters (top two panels)
  over time. The errors correspond to the 3$\sigma$ confidence level for two 
  free parameters. The bottom panels show the count rate in the PCA from 
  2.3-25~keV, which  is a rather 
  accurate match of the 2--10 keV flux (see Tab. \ref{fitres}), and the
  radio flux as measured by the Green Bank Interferometer (GBI)
  (taken from Harrison \etal\ 2000). 
\label{xsp}}
\end{figure}

We find that the spectral photon index and the absorbing column are roughly 
constant throughout
all 11 observations. This is obvious from Fig. \ref{xsp} which
also shows the relevant errors for a two parameter fit (full, correlated 
errors where the error in the absorbing column adds to that in the power law
slope and vice versa).
A notable exception, at first glance, is the spectrum of March 26, 1996
which shows a unique, flat power law photon index of 1.6 (see also 
Tab. \ref{fitres}). However, the best-fit absorbing column for this 
spectrum is as low as 10$^{18}$ cm$^{-2}$ (practically zero), by far less 
than the
foreground absorbing column towards \lsi\ of 
$\sim$8.4$\times$10$^{21}$ cm$^{-2}$ as deduced from HI measurements
(Frail \& Hjellming 1991) or $\sim$6$\times$10$^{21}$ cm$^{-2}$
as deduced from ASCA spectral fitting (Leahy \etal\ 1997). 
This implies that the spectrum
of this date is probably affected by unknown systematic uncertainties
(note that epoch 2 starts already 3 days earlier), and therefore
has to be considered with great care.

This concern is supported by the sequence of ROSAT observations through
an outburst in 1992 which show that the spectrum does not change 
(Taylor \etal\ 1996). In particular, the hardness ratio (their Fig. 3)
is constant except one observation at the radio maximum where it is marginally
harder, suggesting higher absorption (if it were varying absorption).
This is in contrast to the zero-absorption in the single RXTE observation
of March 26 which is nearly exactly at the same radio phase as the
ROSAT outlyer. Assuming that the X-ray/radio outburst behaviour is repeatable
from outburst to outburst, e.g. in 1992 and 1996, then this also
suggests that the RXTE spectrum of March 26 may be unreliable.

\begin{figure}[th]
\vbox{\psfig{figure=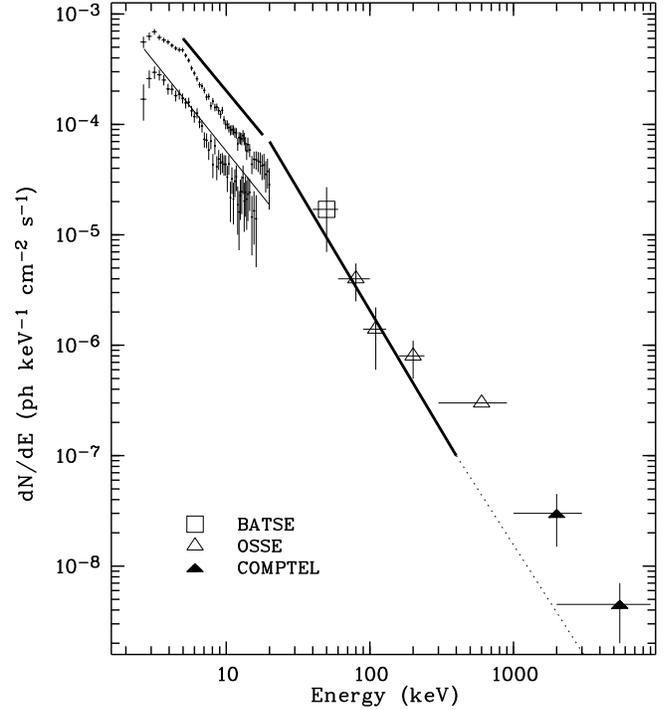,width=8.8cm,%
            bbllx=3.3cm,bblly=8.7cm,bburx=18.7cm,bbury=25.6cm,clip=}}\par
\caption[broadsp]{Observed, not extinction-corrected X-ray
  spectra in the 2--20 keV range observed with RXTE on March 18 
  (small vertical lines; top spectrum)
    and March 10 (bottom spectrum), overplotted over the combined, 
  non-simultaneous BATSE, OSSE and COMPTEL spectra (taken from Apparao 2001).
  The thick line is the expected spectrum due to inverse Compton scattering
  of the optical photons of the Be star by the relativistic electrons
  responsible for the radio emission as predicted by Apparao (2001), with
  the distinctive break in slope around 20 keV. The thin line is the fit 
  of the RXTE spectrum of March 26 which has the flattest power law slope (1.6)
  of all spectra (but see discussion in the text).
  }
\label{broadsp}
\end{figure}

Thus, ignoring for the moment the spectrum of March 26, 1996,
the other 10 spectra are consistent with being constant over a time
interval of 30 days, at mean fit values of $\alpha \sim$ -2.1 and
$N_{\rm H} \sim 4.3 \times 10^{22}$ cm$^{-2}$, and therefore
matching the extension
of the high-energy spectrum up to 10 MeV.
Fig. \ref{broadsp} shows the spectrum of March 10 and 18, respectively, on
top of the OSSE, BATSE and COMPTEL measurements as compiled by Apparao (2001).
The factor 2 and 4, respectively, in normalization difference could be due to
(i) the fact that Apparao (2001) selected observations made around periastron,
   i.e. near the X-ray peak flux, though our selected spectrum of March 18 also
   corresponds to the X-ray peak, 
(ii) the different dates and exposure/integration times of RXTE vs.
  OSSE/BATSE, or
(iii) additional flux by QSO 0241+622 contributing to the OSSE/BATSE flux.
We note that our fluxes are about 25\% larger than those derived by
Harrison \etal\ (2000), most probably because they only included layer 1 data.
The 2--10 keV ASCA fluxes (Leahy \etal\ 1997, Harrison \etal\ 2000) are
5.8$\times$10$^{-12}$ erg cm$^{-2}$ s$^{-1}$ and
4.3$\times$10$^{-12}$ erg cm$^{-2}$ s$^{-1}$,
and thus substantially lower than the RXTE fluxes, thus indicating that there
is nothing wrong with the RXTE fluxes being smaller than the OSSE/BATSE fluxes.

As a side-product of our spectral fitting, we noticed that some RXTE spectra
show a strong residual feature at around 5 keV (see Fig. lower panel of 
\ref{sample}). The centroid energy of this feature is near to the Xe L edge,
causing some worries about being a \lsi\ intrinsic feature. However, the 
fact that the feature 
%(i) the feature is present in some spectra but not in others,
%and (ii) that 
is not present in Crab data taken on March 8, 1996 and
analyzed with the same response matrix created with HEAsoft 5.0.4,
suggest that this feature is neither due to a failure
of the response or background, but related to some source parameter.
This residual seems to be strongest around the X-ray peak, but this
effect could be due to statistics, e.g. better signal-to-noise. It may
be possible that the feature exists at all times, but is only detected
during times of higher X-ray count rate.
At present, its origin remains unexplained, though. We note that 
if it were Fe K emission, it would correspond to a (receding) velocity of
0.2\,c, much faster than the orbital velocity of any of the two binary 
components, or the Be stellar wind.

\section{Conclusion}

Our result of the analysis of the 2--25 keV X-ray spectrum of \lsi\
is twofold: (i) we find that the spectral shape and intensity (normalization)
above around 20 keV is about 2.0-2.3, consistent with the prediction
of Apparao (2001); (ii) We do not find a spectral break in the spectrum
around 20 keV, but the spectrum continues to lower energies, in fact down
to the lowest energies as measurable by RXTE/PCA, as a straight single
power law. This is in contrast to the calculation of Apparao (2001) who
predicted a break in the spectrum around 20 keV. 

Though the RXTE data do not 
extend far beyond 20 keV, the surprising match in slope to the high-energy
spectrum as derived from BATSE, OSSE and COMPTEL implies that variations due
to different observing times and integration times are marginal, unless
the spectrum above 25 keV is not related to \lsi, but a different source.
The good match also implies that the predicted break is unlikely to be at
energies above the RXTE range. Thus, if one wants to maintain the general 
concept of Apparao (2001), based on our result (i), the only remaining 
possibility for a spectral break would be below $\sim$1 keV.

One may therefore ask whether the parameters for the inverse Compton scenario 
(entering Eq. 1) can be tuned such that is compatible with a break below 
$\sim$1 keV.
The energy of the optical photons of the Be star ($\epsilon \sim 10$ eV)
cannot be changed by more than $\sim$20\%. Thus, the factor of $\gax$20
which is needed to shift the break from 20 keV to below 1 keV must come 
through the Lorentz factor 
$\gamma_c$ (Eq. 1), i.e. the product $v_{\rm d} \times r$ of radial
velocity drift $v_{\rm d}$ of the electron bubble and the distance $r$ 
between bubble and Be star would have to be a factor $\gax$4.5 smaller 
(since the break energy depends on $\gamma_c^2$).
Apparao (2001) used $r$ = 3$\times$10$^{12}$ cm (the periastron distance)
and $v_{\rm d}$ = 1000 km/s. While the drift velocity is largely unknown, 
the reduction of the value of $r$ would require that the electron bubble
moves towards the Be star without loosing energy and/or expanding.
Thus, under the assumptions as made by Apparao (2001) it seems unlikely that 
the break can occur below 1 keV.

However, as a more general note we mention
that Apparao (2001) apparently uses the peak radio intensity
to calculate the intensity of the relativistic electrons which then
produce the X-ray spectrum by inverse Compton scattering. However,
the peak radio intensity occurs 10 days  
after the X-ray peak (Fig. \ref{xsp})! Since the electrons
are probably accelerated near periastron, 
they would not only have to survive these 
10 days ($\sim$40\% of the orbital period), but also must not change 
their energy to fulfil the above assumption. Thus, the calculations of 
Apparao (2001) use just a crude approximation, and a time-dependent 
treatment of the electron population and the corresponding X-ray emission
along the orbital phase seems warranted.
It then remains to be seen whether a break $\lax$1 keV would be predicted
(which in any case would be difficult to verify), or whether the idea
of inverse Compton emission has to be abolished at all.

\begin{acknowledgements}
We are extremely grateful to Keith Jahoda (GSFC/USA) for help in the response
matrix generation and for double checking the residual 5 keV feature.
This research has made use of data obtained through the HEASARC,
provided by the NASA/Goddard Space Flight Center.

\end{acknowledgements}

\end{document}